\documentclass[12pt,preprint]{aastex}
\begin{document}
\title{An Interpretation of the  $h\nu_{peak}$ - $E_{iso}$ Correlation  for GRB}
\author{ David Eichler\altaffilmark{1}}
\author{Amir Levinson\altaffilmark{2}}
\altaffiltext{1}{Physics Department, Ben-Gurion University,
Beer-Sheva 84105, Israel; eichler@bgumail.bgu.ac.il}
 \altaffiltext{2}{School of Physics and Astronomy, Tel-Aviv University , Tel Aviv 69978 Israel;}

\begin{abstract}
The $h\nu_{peak}/100 KeV \sim [E_{iso}/10^{52} \rm{erg}]^{1/2}$
correlation  reported recently  by Amati  et al. (2002) and Lamb
et al. (2003) for long GRB's can be interpreted as a viewing angle
effect if the emitting region of the fireball is ring-shaped.
\end{abstract}

\keywords{black hole physics --- gamma-rays: bursts and theory  }

 The  HETE II collaboration (Atteia et al. 2003, Lamb et al. 2003)
has recently announced a pronounced correlation (previously
noticed with a more limited data set by Amati et al. [2002])
between the isotropic equivalent luminosity of GRB's and the
location of the spectral peak at local redshift
 $\nu F(\nu)$. They find

\begin{equation}
h\nu_{peak}/100 KeV \sim [E_{iso}/10^{52} \rm{erg}]^{1/2}.
\label{Enu}
\end{equation}

This relation spans  two orders of magnitude in $\nu_{peak}$, roughly
from 10 KeV to 1  MeV,  and about four orders of magnitude in
$E_{iso}$. The extension to low energies is based on a small
number of X-ray flashes with measurable redshifts. It does not
preclude the possibility that many X-ray flashes derive their low
apparent peak luminosities in part due to cosmological redshift.

Here we consider whether this correlation can be interpreted as
due to viewing angle effects. Off-axis effects have been
considered by previous authors, (e.g. Yamazaki et al. 2003 and
references therein). However, they do not lead to the observed
correlation for a pencil beam.

Suppose  that a source directs emitting material beamed at a fixed
Lorentz factor, $\Gamma$, over a spread of directions {$ \hat
\beta$} relative to the line of sight.  Assuming that each fluid
element emits isotropically in its rest frame a total energy
$E^\prime$, then its contribution to the overall energy emitted
into a solid angle $d\Omega_n$ around the sight line direction
$\hat{n}$, as measured in the Lab frame, is:
$dE(\hat{n},\hat\beta)=(E^\prime D^3/\Gamma)d\Omega_n$, where
$D=1/\Gamma(1-{\bf \beta}\cdot \hat n)$ is the corresponding
Doppler factor. The fluence along the line of sight  is given by
the integral

\begin{equation}
E_{iso}\propto \int\frac
{dE(\hat{n},\hat\beta)}{d\Omega_n} d\Omega_\beta,
\label{Eiso-1}
\end{equation}
where the integral is over the directions $\hat\beta$.

For convenience, let us consider a uniform, axisymmetric jet
of opening angle $\theta_2$, with intrinsic spectral peak at
$\nu^*$, and with a hole of angular size
$\theta_1$ cut out of it.   This
symmetry is strictly for convenience, and it is easily seen that
the scaling presently derived applies to more complicated
geometries.  We take the symmetry axis to lie along the z axis, and
define $\cos\theta_\beta=\hat{\beta}\cdot\hat{z}$, and $\cos\theta_n=\hat{n}\cdot\hat{z}$.
Then $\hat{n}\cdot\hat{\beta}=\cos\theta_\beta\cos\theta_n +\sin\theta_\beta\sin\theta_n\cos\phi$,
where $\phi$ is the azimutal angle, and eq. (\ref{Eiso-1}) reduces to:

\begin{equation}
E_{iso}\propto \int_{\theta_{1}}^{\theta_{2}} \int_{0}^{2\pi}{
\frac{\sin\theta_\beta d\theta_\beta d\phi}{[1-\beta(\cos\theta_\beta\cos\theta_n
+\sin\theta_\beta\sin\theta_n\cos\phi)]^3}}.
\label{Eiso}
\end{equation}

The observed spectral peak is located at
$\nu_{peak}=\nu^\star/\Gamma[1-\beta\cos(\theta_n-\theta_2)]$
for sight lines outside the jet (that is; $\theta_n>\theta_2$) and
$\nu_{peak}=\nu^\star/\Gamma[1-\beta\cos(\theta_1-\theta_n)]$ for sight lines inside
the hole ($\theta_n<\theta_1$).

It is straightforward to see that when the line of sight is
extremely close to the boundary of the jet (but still large
compared to $1/\Gamma$), the only scale in the integral is the
angular separation $|\theta_n-\theta_\beta|$, and eq. (\ref{Eiso})
yields to the lowest order $E_{iso}\propto
(\nu_{peak}/\nu^\star)^2$, as reported by the observers.
Similarly, for a "fractal"-like jet where the geometry and
statistics of the patches are such that the solid angle of the
emitting material is comparable to the distance to the nearest
patch, the same relation would hold. The fractal dimension of the
emitting surface should be slightly less than 2. To examine the
relation between $E_{iso}$ and $\nu_{peak}$ over a broad range of
viewing angles, we integrated eq. (\ref{Eiso}) numerically.  The
results obtained for the region inside the jet are shown in fig.
1, and outside the jet in fig. 2.  The annulus is centered around
$\theta_0=0.1$ in this example and its angular width is
$\Delta\theta<\theta_0$.  The corresponding range of viewing
angles extends from $\theta_n=0$ to about $\theta_n=0.4$.  Note
that the $E_{iso}$ - $\nu_{peak}$ relations inside and outside the 
jet are not identical.  This introduces some scatter in the correlation,
however, we find this scatter to be rather small.

A fair example, we believe, is a thick annulus, whose overall
outer edge is only several times large than its thickness, which
in turn is not larger, by a very large factor, than $1/\Gamma$. It
is clear that this geometry increases the probability, relative to
that of a solid jet, of the line of sight being near the beam
without being in it.  This in turn would raise the number of X-ray
flashes relative to the number of hard GRB's. Yet the significant
contribution to the line of sight comes from a solid angle roughly
proportional to that of the largest neighborhood surrounding the
line of sight that does not touch the emitting parts of the jet.
Invoking fractal-like corrugation in the boundary of the emitting
region enhances this probability even further, and raises the
number of off-axis to "on-axis" viewing angles. A "patchy" jet
(Nakar and Piran 2001) could have the same effect. The required
condition is that the solid angle of the (soft) $\gamma$-ray-quiet
regions emitting regions  is comparable to that of the emitting
regions and that the scale of variation is of order $1/\Gamma$.

Why would the GRB fireball be shaped like a thick annulus? One
plausible reason  is that the soft gamma ray emission comes mainly
on that part of the jet that is baryon loaded, and that the baryon
loading comes from the periphery.  In the particular case of
baryon loading by neutron leakage from the walls of a confining
wind or stellar envelope (Levinson and Eichler 2003), the neutrons
are quickly charged by collisions near the walls, before they can
penetrate to the center. It was shown in the above reference that
the annular region that is significantly loaded can have a solid
angle that is not too much less than the inner hollow region.

Why would the Lorentz factor be connected to the angular thickness
of the jet? One possibility is that the  material at the walls
that collimate the GRB fireball is itself accelerated to a Lorentz
factor that is of order $1/\theta_o$ (Eichler 2003) where
$\theta_o$ is the opening angle of the jet as it emerges from the
host star. $\gamma$-rays emitted or reflected by such material
would be beamed into a cone within $1/\Gamma \sim \theta_o$ of the
emitting/scattering material, and this would establish the
thickness of the annulus.  Of course, $\Gamma$, whatever
establishes its value, sets the smallest scale of the patchy
emission, so the above scenario need not be the only one that can
work in the present context.

We now consider the question of whether all this is consistent
with the observed distribution of $\nu_{peak}$.  Let $\nu_{obs}(\mu)=
\nu_{peak}(\mu)/(1+z)$, where $\mu=\cos\theta$, be the peak frequency
observed at angle $\theta$ to the jet axis at redshift $z$.
The number of bursts seen per unit time per $\ln \nu_{obs}$
from cosmological redshifts between z and z+dz is given by

\begin{equation}
\frac{dN(z,\nu_{obs})}{d\ln\nu_{obs}} = (1+z)^{-1}
\left(\frac{\partial \ln \nu_{obs}}{\partial \mu}\right)^{-1}
f(z)r(z)^2dr(z) \label{dN}
\end{equation}
where the factor $(1+z)^{-1}$ expresses the fact that the burst
rate for a galaxy as seen by the  distant observer is slowed by
$(1+z)$, $r(z)$ is the comoving coordinate of a GRB source that is
just now being detected, and $f(z)$ is the proper rate of GRB host
formation within an average galaxy.  For the annular jet model
discussed above there are three regimes: 1) within the confines of
the  annulus $\theta_1 \le \theta \le  \theta_2$  (i.e., head-on
observers), 2) inside the hole in the annulus $\theta \le
\theta_1$ and 3) outside the annulus $\theta\ge \theta_2$.

For head-on observers, $\nu_{peak}=2\Gamma\nu^\star$ is fixed and the change in $\nu_{obs}$ is
solely due to redshift effects.  In this regime we
obtain $\left(\partial \ln \nu_{obs}/\partial \mu\right)^{-1}=
\sin\theta_0\Delta\theta(1+z_\nu)\delta(z-z_\nu)$, where
$z_\nu=(2\Gamma\nu^\star/\nu_{obs})-1$, and a total rate per
$\ln \nu_{obs}$ of $(dN/d\ln\nu_{obs})_{head-on}=
\sin\theta_0\Delta\theta [r(z_\nu)]^2(dr/dz)f(z_\nu)$.

Outside the annulus
$\left(\partial \ln \nu_{obs}/\partial \mu\right)^{-1}=
(1+z)^{-1}(\nu^\star/\Gamma\nu_{obs})(\sin\theta/\sin\Delta)$, where
$\Delta=\theta_1 -\theta$ inside the hole and $\Delta=\theta -\theta_2$
outside the jet.  The total rate per $\ln\nu_{obs}$ in this regime is then
obtained by integrating the right  hand side of (\ref{dN}) over $z$ from $z=0$
to $z=z_{max}(\nu_{obs})$, with $z_{max}(\nu_{obs})=min\{z_0,z_\nu\}$, where $z_\nu$
is defined above and, for a detector of
a threshold fluence ${\cal F}_{th}$, $z_0$ is given implicitly
through $d_{L}(z_{0})=[E_{iso}(\nu_{peak})/4\pi {\cal F}_{th}]^{1/2}$, with $d_L$
being the luminosity distance.
The net distribution $dN/d\ln\nu_{obs}$ is the sum of the distributions in these
three regions.

The resultant rate distribution $dN/d\ln\nu_{obs}$, calculated
using a $\Lambda$CDM cosmology, is shown in figs. 3-5.  The
parameter $d_L^\star$ denotes the maximum luminosity distance (in
units of $R_0$) below which a source observed head on is above
detection limit.  To be precise $d_L^\star=(E^\star/4\pi
R_0^2{\cal F}_{th})^{1/2} =10^{2} [(E^\star/2\times10^{54}\ {\rm
erg})/({\cal F}_{th}/10^{-7}\ {\rm erg\ cm^{-2}})]^{1/2}$, where
$E^\star=E_{iso}(\nu_{peak}=2\Gamma\nu^\star)$ is the isotropic
equivalent energy that corresponds to viewing angles within the
annulus, and where a Hubble constant $h= 0.75$ has been adopted.
For simplicity we assume that the threshold fluence is energy
independent. Fig. 3 exhibits the dependence of the rate
distribution on the angular width of the jet for the case $f(z)=1$
(no evolution).  As seen, there are quite generally two peaks. The
first peak at $\nu_{obs}\simeq 0.25 (2\Gamma\nu^\star)$ is
contributed by viewing angles within the annulus (that is, the
peak of $(dN/d\ln\nu_{obs})_{head-on}$). The location of this peak
corresponds merely to the redshift at which the volume $r^2dr$ is
maximum.  The second peak at $\nu_{obs}\simeq 0.025
(2\Gamma\nu^\star)$ is contributed by the distributions outside
the annulus.  The sharp decline in the number of bursts below the
second peak is due to the rapid decrease in the volume occupied by
those bursts that exceed detection limit. The location of this
peak depends on the parameter $d_L^\star$ defined above, as seen
in fig. 4. The effect of redshift evolution is examined in fig. 5.
The curve labelled SF2 was computed using $f(z)=(1+z)^3$ for
$z\le2$ and $f(z)=27$ for $z>2$.

We therefore conclude that the interpretation of eq. (\ref{Enu}) as
being due to the viewing angle is compatible with X-ray flashes
being quite common. It is important to include the contribution of
the cosmological red-shift in determining the observed spectral
peak.

In summary, we have shown that off-axis viewing of an annular jet
leads to the correlation that the isotropic-equivalent fluence
$E_{iso}$ is roughly  proportional to the square of the spectral
peak frequency $\nu_{peak}$ as noticed  by various observers. The
combination of off-axis viewing and cosmological redshift "smear",
a single value of $E_{iso}$ over  more than an order of magnitude
- roughly from 0.025 to 0.3 of the intrinsic spectral peak
$\nu^*$ - over which the number density of sources per unit
logarithm in the apparent spectral peak $\nu_{obs}$, is roughly
constant when the thickness of the  annulus is about $5/\Gamma$.
Thus, if $\nu^* = 1 MeV$, $\theta_o \sim 0.1$, $\Gamma \sim 10^2$,
$\Delta \theta \sim 0.05$, most GRB spectral peaks would be
expected to lie between 30 and 300 KeV. In the likely event that
there is some intrinsic spread in $\nu^\star$,  the range of
$\nu_{obs}$ is enhanced accordingly. Moreover, we have
conservatively assumed in the above an energy independent fluence
threshold; if the detection threshold for X-ray flashes is lower
in energy fluence, e.g. if it is constant in photon number, the
detectable X-ray flashes would populate a region of even lower
$\nu_{obs}$.

If the annulus  is reasonably thick, then it subtends a
significant fraction of the solid angle subtended by its outer
cone and non-simple geometry should not disrupt the inverse
correlation between $E_{iso}$ and jet opening angle (Frail et al.
2001). Note that there is an intrinsic spread in $E_{iso}$ even
when the total GRB energy is without any spread, this being  due
merely to the spread in opening angle, and in the present context
this implies at least some scatter in the Amati et al.
correlation. However, the correlation between $\Delta \theta$ and
$\theta_o$ is still uncertain in the proposed picture.

A post-break switch-on of afterglow would provide some
observational support for this picture, as it could be interpreted
as the earlier parts of the afterglow being beamed away from the
observer. On the other hand, it is conceivable that for an
observer looking down the hole in the annulus, the afterglow could
nonetheless be strong if the jet in this inner region is comprised
of "virgin" Poynting flux which would produce afterglow but little
soft gamma radiation.

We thank Drs. D. Lamb, C. Graziani  and Y. Lyubarsky for helpful
discussions. This research was supported by the Israel-US
Binational Science Foundation, an Israel Science Foundation Center
of Excellence Award, and the Arnow Chair of Theoretical Physics.

\clearpage
\newpage
\begin{figure}
\plotone{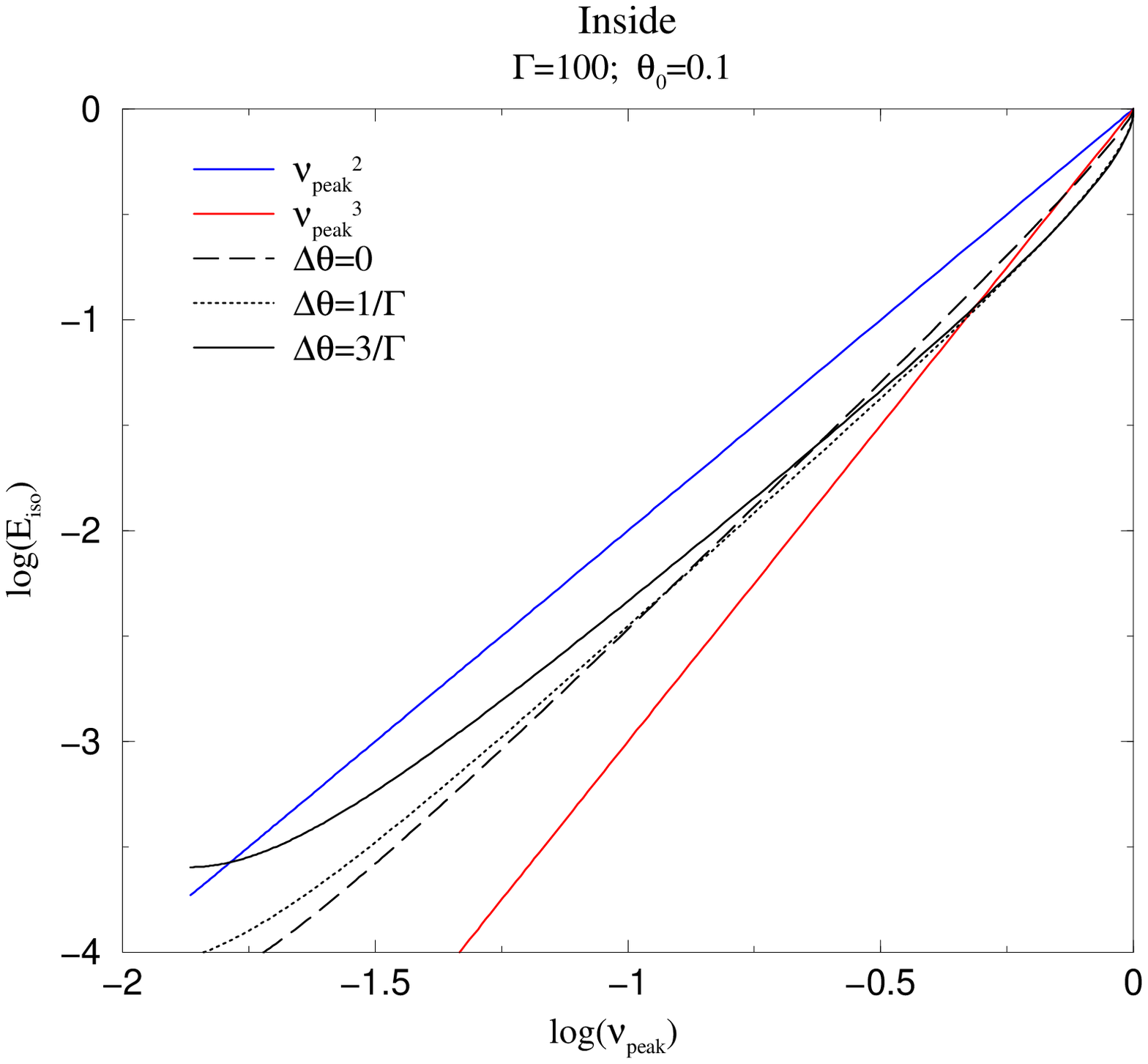}
\caption{The relation between the isotropic equivalent energy $E_{iso}$ and
SED peak frequency, $\nu_{peak}$, calculated inside a hollow jet of
angular width $\Delta\theta$. The velocity vectors of the emitting material
lie in the range between $\theta_1=\theta_0-\Delta\theta/2$ and
$\theta_2=\theta_0+\Delta\theta/2$.  The range of viewing angles in this
plot is $0\le\theta_n\le\theta_1$.}
\end{figure}

\begin{figure}
\plotone{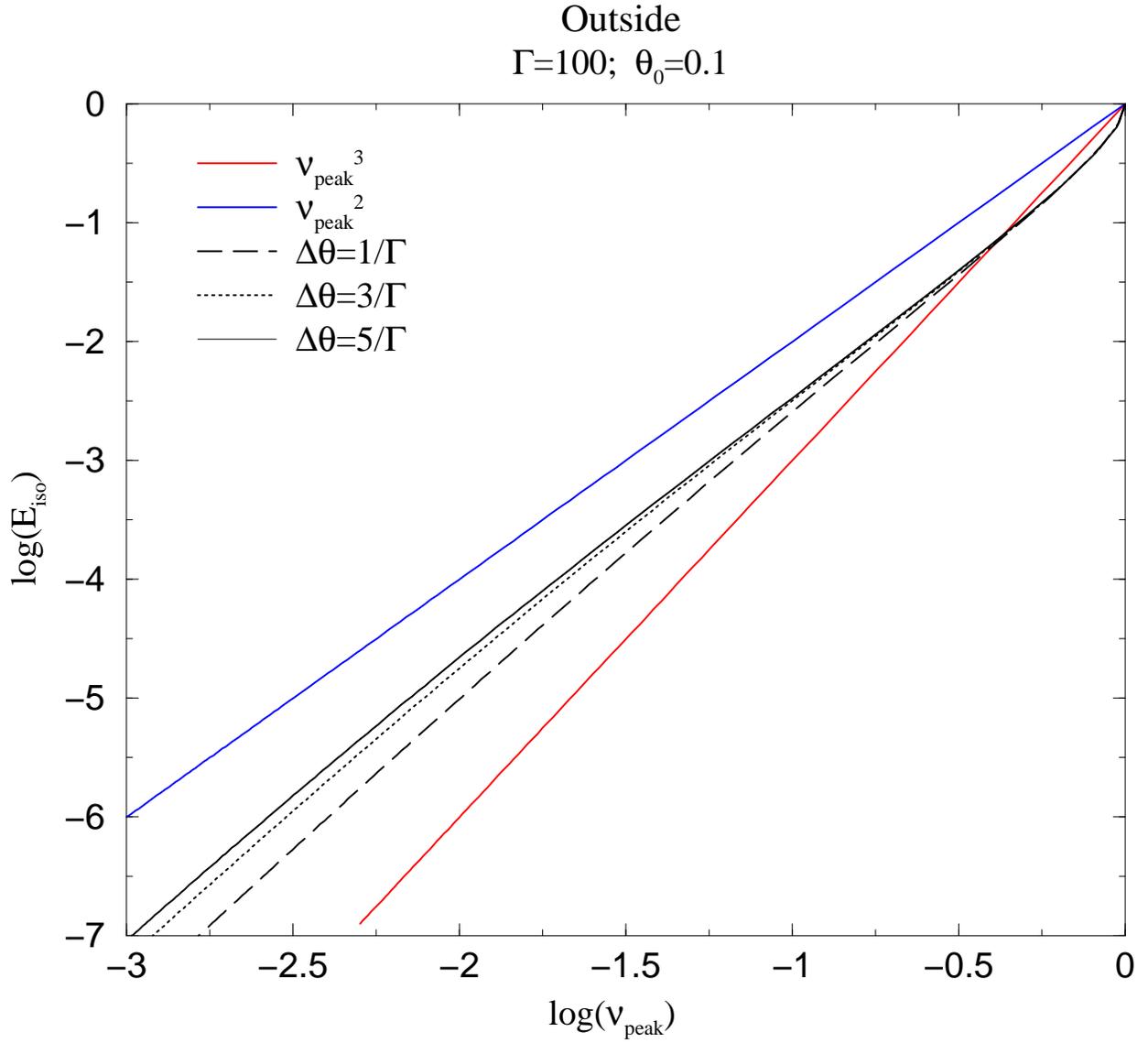}
\caption{Same as fig. 1 but for viewing angles outside the jet;
$0.4\ge\theta_n\ge\theta_2$. }
\end{figure}

\begin{figure}[hb]
\plotone{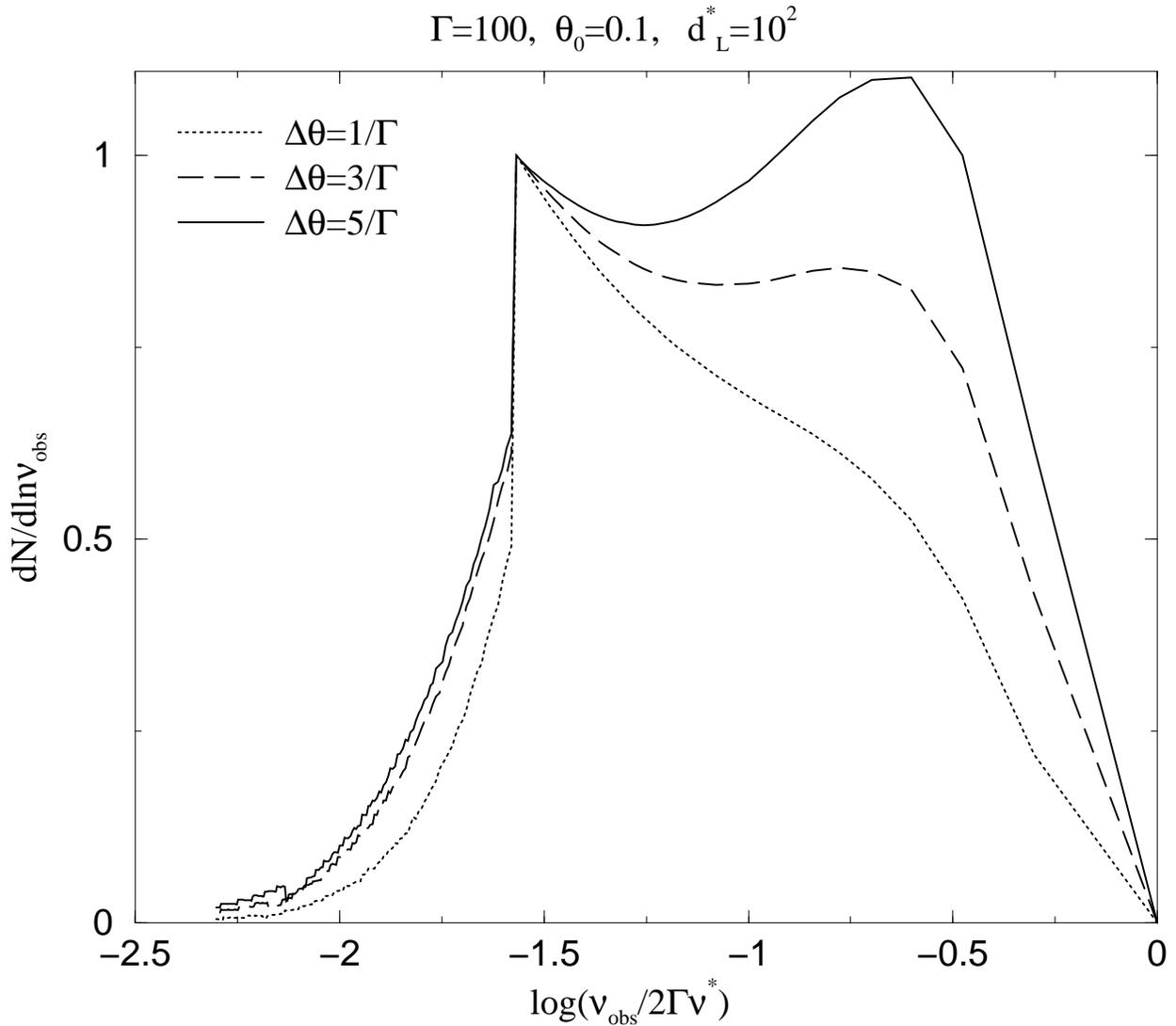}
\caption{Number of bursts seen per unit time per logarithmic
interval of observed peak frequency versus log of the normalized
peak frequency, obtained for the annular jet model discussed in
the text.  The parameters $\theta_0$, $\Delta\theta$, and $\Gamma$
are, respectively, the opening angle of the jet, its angular
width, and the bulk Lorentz factor of the emitting material.}
\end{figure}

\begin{figure}[hb]
\plotone{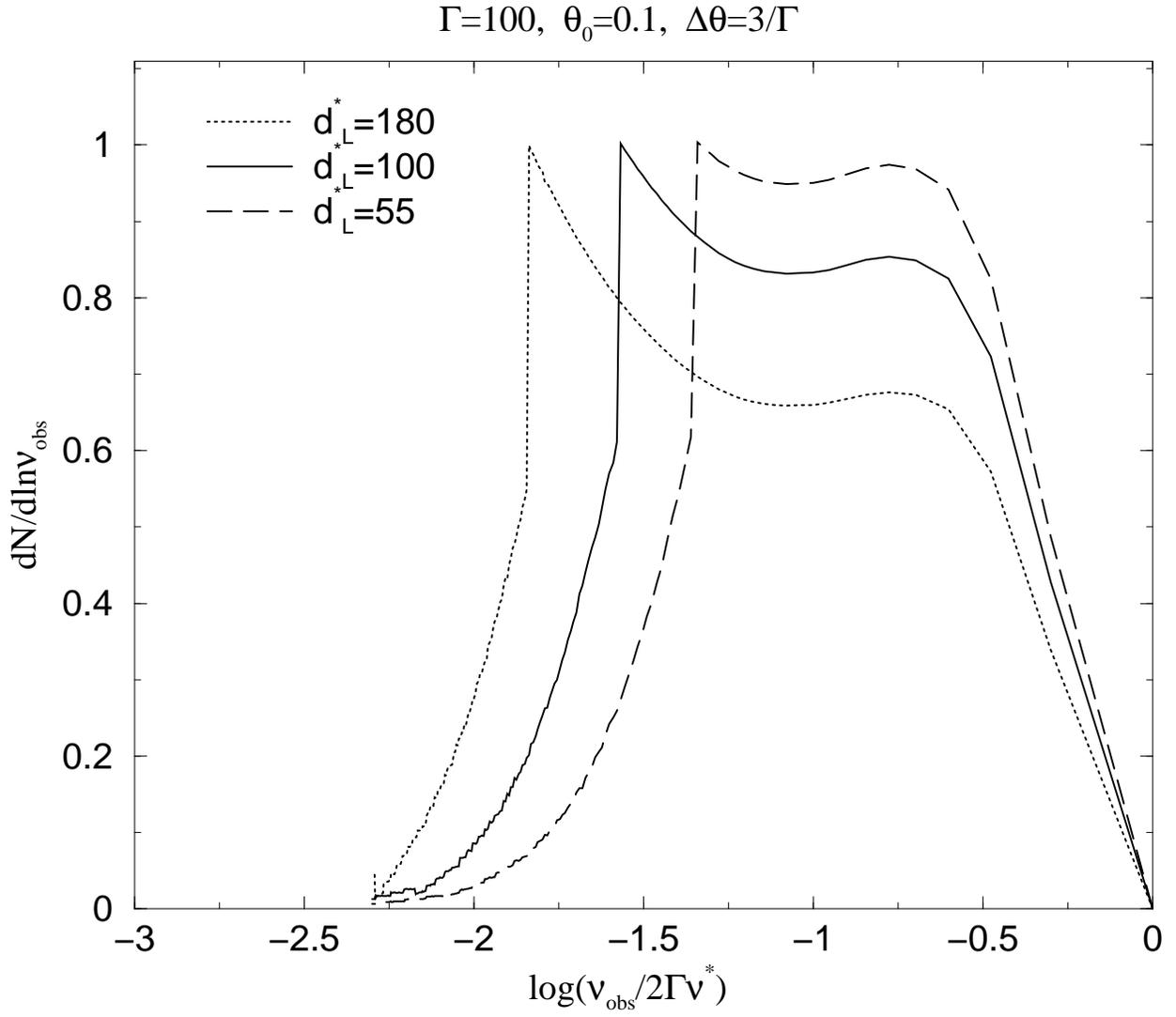}
\caption{Same as fig. 3 but for different values of
$d_L^\star$}
\end{figure}

\begin{figure}[hb]
\plotone{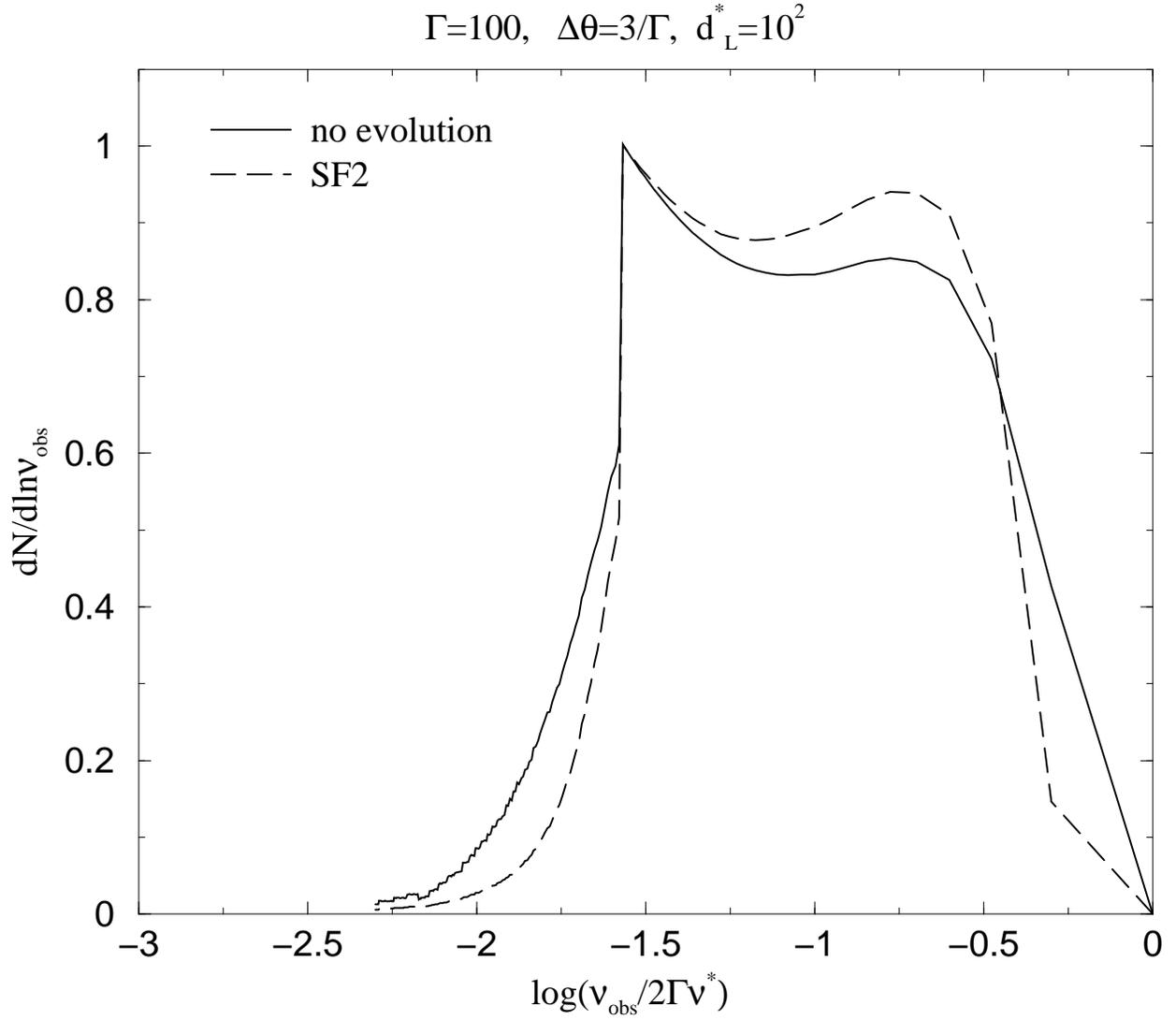}
\caption{Effect of redshift evolution on the rate distribution of bursts (see text
for discussion).}
\end{figure}

\end{document}